\documentclass[11pt]{article}
\usepackage{graphicx,layout,wrapfig}

\begin{document}

\fontfamily{cmss}
\selectfont

\begin{figure}[ht]
\includegraphics[width=16cm]{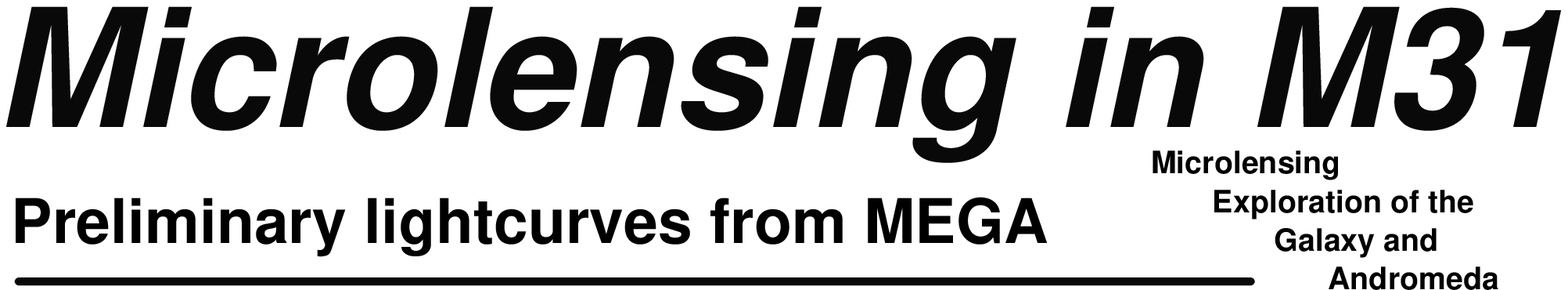}
\end{figure}

\begin{centering}
{\large Jelte T.A. de Jong$^1$, Penny D. Sackett$^1$, Konrad H. 
Kuijken$^1$, Robert R. Uglesich$^2$,\\Arlin P.S. Crotts$^3$, Will J. Sutherland$^4$}\\
{\normalsize $^1$Kapteyn Astronomical Institute, University of Groningen\\ $^2$Steward
Observatory, University of Arizona\\ $^3$Department of Astronomy,
Columbia University\\ $^4$Department of Physics, University of Oxford}\\
\end{centering}

\vskip 0.9cm

\begin{abstract}
\noindent
One of the possible astrophysical solutions to the galactic dark
matter problem is the presence of a significant amount of ``dark''
compact objects (MACHOs) in galactic dark matter halos.
MEGA (Microlensing Exploration of the Galaxy and Andromeda) tries to
find proof for or against the presence of compact objects in the halo
of the Andromeda galaxy (M31) by looking for the microlensing
signature that would be induced by these objects.
The lightcurves presented here are preliminary and based on
observations of M31 with the Isaac Newton Telescope (INT) on La Palma
during the second half of 1999.
\end{abstract}

\vskip 0.9cm

\noindent
{\huge 1~~Introduction}\\

During the past decade microlensing has proven to be a useful probe of
dark matter in the Milky Way. If the dark halo of our galaxy would 
(partly) consist of massive compact objects (MACHOs) then they should
reveal their presence by magnifying background sources when passing
exactly in front of them, as first suggested by Paczynski
(1986). Recent surveys in the direction of the Magellanic Clouds show
that up to 20\% of the halo mass might be in compact objects of $\sim
0.5 M_{\odot}$. However, the uncertainty is still rather large,
because both the nature and the location of the lenses is not clear.\\

\vskip 0.5cm

\begin{wrapfigure}[16]{r}{5.5cm}
\includegraphics[angle=270,width=5cm,clip=]{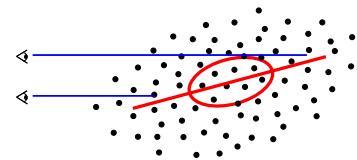}
\fontfamily{cmss}
\selectfont
\caption{
{\small If the Andromeda galaxy is surrounded by a
halo of dark compact objects, the sightline to the far side
of the disk goes through a much larger part of the halo than the
sightline to the near side. Thus, the probability of microlensing is
much larger towards the far side.}}
\label{cartoon}
\end{wrapfigure}
\noindent
{\huge 2~~The MEGA project}\\

In the case of the nearby Andromeda galaxy (M31), microlensing can
be used to unambiguously establish the presence or absence of a halo
made up of MACHOs (Crotts, 1992).
Because of the high density of background stars, the microlensing
optical depth $\tau$ towards M31 is expected to be an order of
magnitude larger than towards the Magellanic Clouds. A survey of small
fields in M31 has already produced several microlensing candidates at
roughly the expected rate (Crotts \& Tomaney, 1996).

\begin{figure}[t!]
\begin{center}
\includegraphics[angle=270,width=13.7cm,clip=]{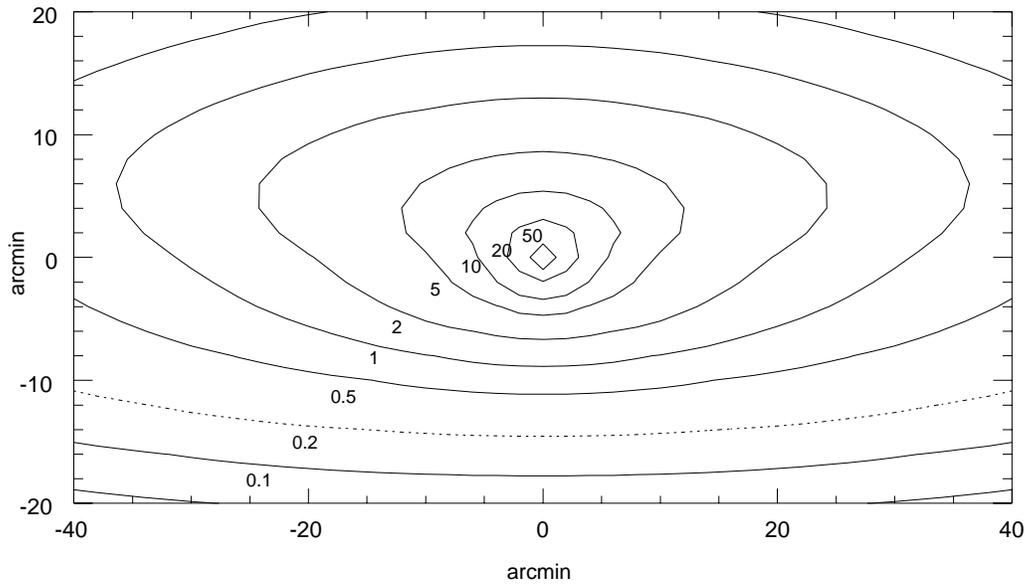}
\fontfamily{cmss}
\selectfont
\caption{
{\small An 80x40 arcmin plot of the center of M31 (major axis is
horizontal, minor axis vertical) showing contours of
the predicted event rate (year$^{-1}$ arcmin$^{-1}$) for bulge and
halo microlensing.
This model assumes an unflattened halo with a 5 kpc core radius,
predicting over 100 events during an M31 observing season (Aug-Jan)
for an experimental setup similar to that of MEGA. Taken from Crotts
et al. (2000).}}
\label{rates}
\end{center}
\end{figure}

Furthermore, the high inclination of M31 provides an observer
on earth with many different sightlines through the M31 halo
(illustrated in figure~\ref{cartoon}). This 
means that the microlensing event rate due to halo lenses should show
a strong asymmetry. 
The Microlensing Exploration of the Galaxy and
Andromeda (MEGA) tries to find proof for or against the presence of
compact objects in the halo of M31 by mapping the distribution of
microlensing events and looking for a gradient from the near to the
far side of the disk. An example of how the predicted microlensing
event rates may vary over the galaxy for a specific halo model and a
survey setup similar to that of MEGA is shown in figure~\ref{rates}.

To detect microlensing events in M31 two large fields ($\sim 0.3~
deg^2$ each) are being monitored for three years. 1999 was the first
year of observing and the survey will continue at least until January
2002.
Such a long-term monitoring period is needed to detect enough events to
convincingly identify the gradient that would be caused by a MACHO
halo. With a large enough number 
\begin{wrapfigure}[21]{r}{7cm}
\includegraphics[angle=270,width=6.9cm]{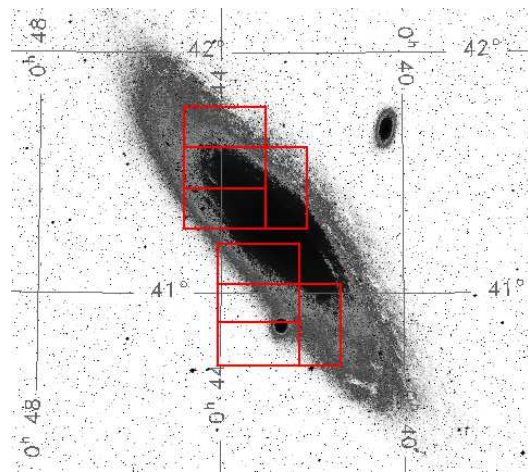}
\fontfamily{cmss}
\selectfont
\caption{
{\small The two INT Wide Field Camera (WFC) fields that are monitored
by MEGA, each covering an area of approximately 0.3 $deg^2$. 
Both fields are also monitored with wide field cameras on the 1.3m and 
2.4m telescopes of the MDM observatory and the KPNO 4m telescope.}}
\label{fields}
\end{wrapfigure}
of events it might even be possible
to constrain some of the shape parameters of the halo, if it is there.

A long time baseline is also needed to distinguish microlensing events
from certain types of long period variable stars. Color information is
also very useful for distinguishing between variable stars and
microlensing, since gravitational lensing is monochromatic, while stellar
variability generally implies color changes. For this microlensing
survey M31 is observed in three bands, namely g', r' and i'.

Several telescopes are used to
monitor both our fields, namely the Isaac Newton Telescope on La
Palma, the 1.3m and 2.4m telescopes of the MDM observatory near
Tucson, Arizona, and the KPNO 4m telescope at Kitt Peak. With these
telescopes we have good time coverage during the whole M31 observing
season, running from August to January. 
The layout of the INT fields is shown in figure~\ref{fields}.
In section 4 we present some preliminary lightcurves based on data taken with
the Isaac Newton Telescope (INT) during the second half of 1999.\\

%\vskip 0.5cm

\noindent
{\huge 3~~Our method}\\

Because of the high stellar density in M31 and its large distance, the
background source stars are usually only resolved while being lensed
and sufficiently magnified. Pilot surveys have already shown that it
is nevertheless possible to find microlensing events (Crotts \&
Tomaney, 1996; Ansari et al., 1999). We use difference image
photometry (Tomaney \& Crotts, 1996) to detect and photometer variable
and often unresolved sources in M31.

\begin{figure}[t]
\begin{center}
\includegraphics[width=4.9cm,clip=]{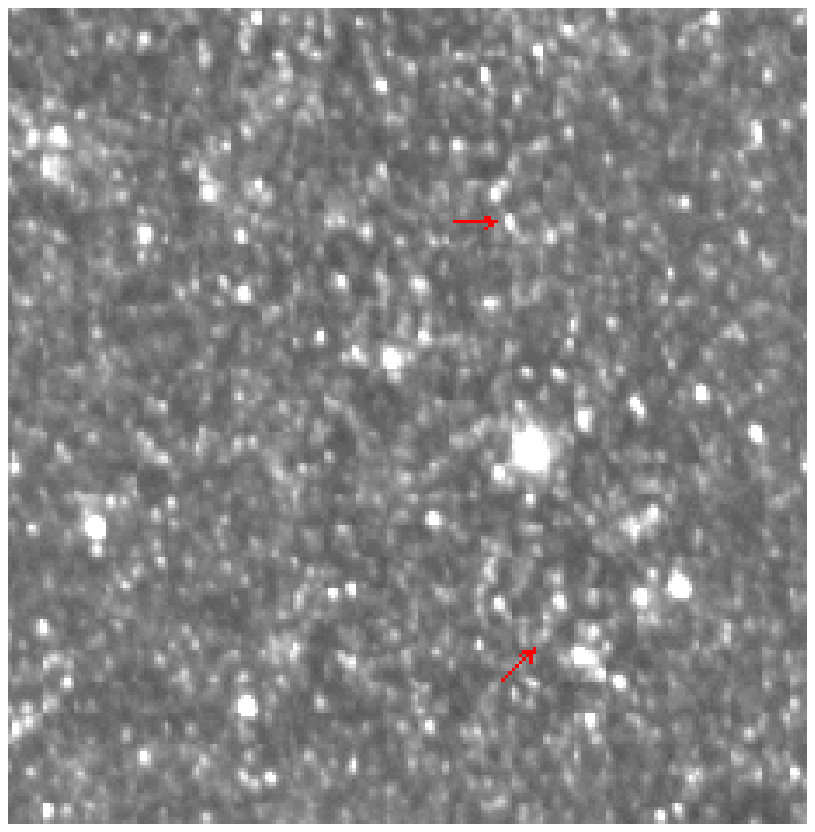}
\includegraphics[width=4.9cm,clip=]{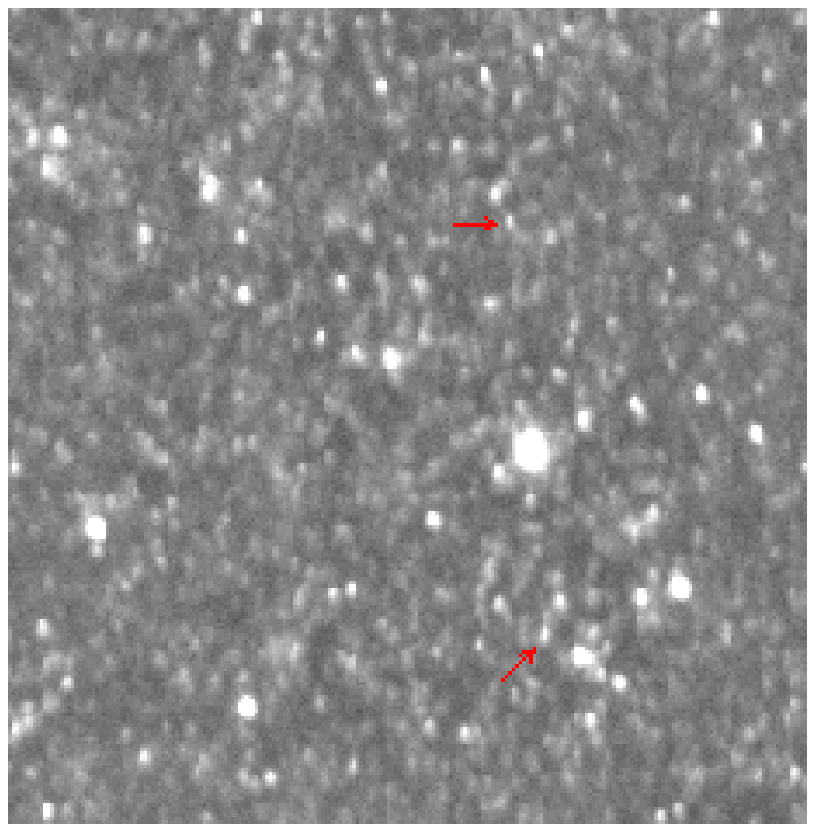}
\includegraphics[width=4.9cm,clip=]{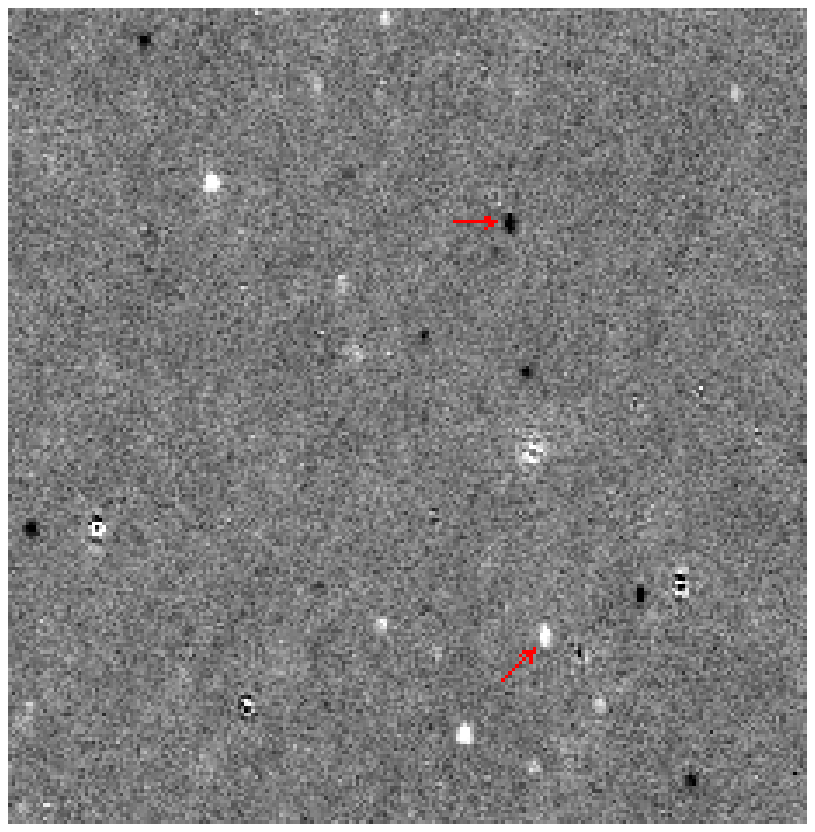}
\fontfamily{cmss}
\selectfont
\caption{
{\small An example of image subtraction with INT WFC images. The left
panel shows part of a high S/N reference image. The middle panel shows
the same part of a single night image. After psf-matching the images
are subtracted, resulting in the difference image shown in the right
panel. Variable sources show up as positive and negative residuals on
a background dominated by shot noise.}}
\label{dipexample}
\end{center}
\end{figure}

The standard data reduction for the INT Wide Field Camera (WFC) images
is performed in IRAF. A high signal-to-noise reference image is then 
created from some of the best seeing frames taken over a whole observing
season (Aug-Jan). Nightly images are subtracted from this reference
image, resulting in difference images in which variable sources remain
as positive or negative residuals. Before subtracting two images the
point spread functions (PSFs) are matched by degrading the best seeing
frame (usually the reference frame) to the same seeing as the worst
seeing frame.
The variable sources can be detected easily in the difference images,
after which aperture photometry is used to construct lightcurves.
An example of the image subtraction method is shown in 
figure~\ref{dipexample}.\\

\vskip 0.5cm

\noindent
{\huge 4~~Some lightcurves}\\

\begin{figure}[ht]
\begin{center}
\includegraphics[width=14cm,clip=]{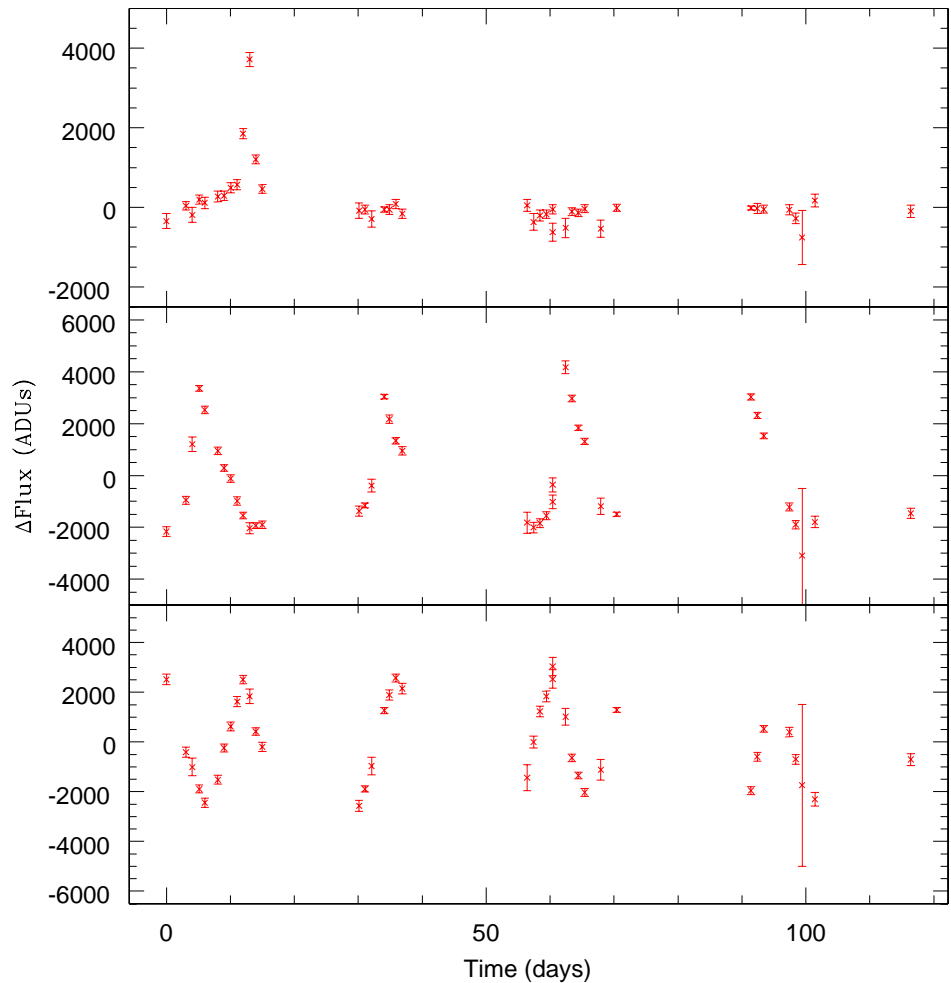}
\fontfamily{cmss}
\selectfont
\caption{
{\small Three r' lightcurves obtained from MEGA INT images taken
during Aug-Nov 1999 (day 0 corresponds to Julian Date
2451393.59405). The upper panel shows the first microlensing event
candidate discovered. The lower two panels show examples of
variable star lightcurves from the same field as the microlensing
candidate. Aperture photometry was performed on single night
difference images. The $\Delta$Flux is relative to the reference
image.}}
\label{lightcurves}
\end{center}
\end{figure}

Three preliminary lightcurves obtained from MEGA INT fields are
presented in figure~\ref{lightcurves}. These lightcurves were measured
from r'-band images during the months August through November of
1999. Single points in the lightcurves represent one night, where the
total exposure time per night is typically 10 or 20 minutes. The
fluxes are measured relative to a high S/N reference image.

The upper lightcurve shows the first microlensing candidate event
that has been discovered in this dataset (Auriere et al. 2001). The
short duration peak that occurs around August 15th can be fit well by
standard microlensing models. Auriere et al. (2001) have almost
certainly identified the source star on Hubble Space Telescope
archival images and infer an Einstein crossing time of $t_E$ = 10.4
days and a maximum magnification of $A_{max} \sim$ 18. 
The event could be caused by an object
of 0.06 $M_\odot$ in the halo of M31, but other possibilities are
lensing by a star in M31 or by an object in the Milky Way.

Of all the variable objects that are detected, only very few will be
microlensing events, the majority being variable stars. Two
variable star lightcurves are shown in the lower two panels of
figure~\ref{lightcurves} as examples. These two variable stars
have relatively short periods of about 25 and 30 days and are easily
distinguished from microlensing lightcurves. However, some types of
long-period Mira variables have lightcurves that can closely resemble
microlensing lightcurves. For this reason, long baselines are needed to
exclude periodic variability in microlensing candidates.\\

\vskip 0.4cm

\noindent
{\huge References}\\

\noindent
Ansari, R., et al., 1999, A\&A, 344, L49\\
Auri\`ere, M., et al., 2001, ApJ, submitted (astro-ph/0102080)\\
Crotts, A.P.S., 1992, ApJ, 399, L43\\
Crotts, A.P.S. \& Tomaney, A.B., 1996, ApJ, 473, L87\\
Crotts, A.P.S. et al., 2000, to appear in ``Microlensing 2000: A New
Era of Microlensing\\ 
\indent Astrophysics'' ASP Conf. Series, vol 239 (astro-ph/0006282)\\
Paczynski, B., 1986, ApJ, 304, 1\\
Tomaney, A.B. \& Crotts, A.P.S., 1996, AJ, 112, 2872\\

\end{document}